\documentclass[twocolumn,prl,amsmath,amssymb,floatfix]{revtex4}
\usepackage{graphicx}

\begin{document}

\title{Analytical Characterization of Oscillon Energy and
Lifetime}

\author{Marcelo Gleiser}
\email{gleiser@dartmouth.edu}

\author{David Sicilia}
\email{davidovich@dartmouth.edu}

\affiliation{Department of Physics and Astronomy, Dartmouth College,
Hanover, NH 03755, USA}

\date{\today}

\begin{abstract}
We develop an analytical procedure to compute all relevant physical properties
of scalar field oscillons in models with quartic polynomial potentials: energy,
radius, frequency, core-amplitude, and lifetime. We compare our predictions to
numerical simulations of models with symmetric and asymmetric double well
potentials in three spatial dimensions, obtaining excellent agreement. We also
explain why oscillons have not been seen to decay in two spatial dimensions.
\end{abstract}

\maketitle

\section{Introduction}
It is well known that a real scalar field under the influence of certain
nonlinear potentials supports long-lived, spatially-localized,
time-dependent configurations known as oscillons \cite{bogol,gleiser}. Even
though there are no topological or nontopological conserved charges associated
with such objects, they exhibit remarkably long lifetimes, far
exceeding naive estimates from linear approximations to the theory
\cite{copeland}.  This property has prompted much interest in studying oscillons
in more detail in two \cite{2dosc,hindmarsh1} and three \cite{3dosc} spatial
dimensions. Structures with similar qualitative behavior have been observed in
many different physical systems, from vibrating grains \cite{oscvib} to acoustic
oscillations in the sun \cite{sun}.

In the past two years, it has been shown that oscillons are not limited to real
scalar field models; they have been found in 2d Abelian-Higgs models
\cite{thoru1}, and in the Standard Model \cite{farhi}. They may also
play a role in cosmology \cite{osccos}, possibly having an impact on the
dynamics of
inflationary reheating, in symmetry-breaking transitions, and in false vacuum
decay \cite{oscnuc}. In $1d$ de Sitter models, oscillons not only form from
near-thermal initial conditions, but survive to retain approximately 50\% of the
total energy of the universe \cite{graham_cos}. 

So far, most oscillon properties, such as their energy, radius, and lifetime,
have been obtained numerically. It has been shown
that they have a near-constant energy and a near-periodic motion with
frequency just below the mass of vacuum excitations \cite{hindmarsh1,saffian}.
In spite of this progress, it is fair to say that a proper analytical
understanding of oscillons is still lacking.

In the present work, we will derive all physically relevant properties of scalar
oscillons in quartic polynomial potentials. Starting from a simple {\it ansatz}
for their profile, we develop a \textit{nonperturbative} analytical procedure to
obtain their energy, lifetime, and minimum and maximum values
of their frequency, radius, and core amplitude. We compare our analytical
results to numerical simulations obtaining excellent agreement.

\section{General Framework}
We begin with the Lagrangian for a $d$-dimensional, spherically-symmetric real
scalar field in flat spacetime,
\begin{equation}
\label{lagrangian1}
L = c_d\int r^{(d-1)}dr \left( \frac{1}{2}\dot\phi^2 - 
\frac{1}{2}\left (\frac{\partial\phi}{\partial r}\right )^2
-V(\phi)\right),
\end{equation}
\noindent
where a dot denotes a time derivative, and $c_d=2\pi^{d/2}/\Gamma(d/2)$.
Inspired by the numerical solutions, which show that an oscillon is well
approximated by a Gaussian undergoing near-periodic motion in its amplitude and
with an effective radius oscillating with small amplitude about a mean value
\cite{gleiser, copeland}, we approximate the oscillon solution as
\begin{equation}
\label{ansatz}
\phi(t,r) = A(t)\exp \left(-\frac{r^2}{R^2}\right) + \phi_v,
\end{equation}
where $A(t)\equiv\phi(r=0,t)-\phi_v$ is the displacement from the vacuum,
$\phi_v$. This approximation works particularly well in $d=2$ and $d=3$,
although it clearly
fails to reproduce the oscillon's large $r$ behavior, where
$\phi(r)\sim \exp[-mr]$.

Since oscillons have been observed in a diversity of quartic polynomial
potentials, we write $V(\phi)$ as \cite{dosc}
\begin{equation}
\label{potential}
V(\phi) = \sum_{j=1}^{4} \frac{g_j}{j!}\phi^j - V(\phi_v)~,
\end{equation}
\noindent
where the $g_j$'s are constants and $V(\phi_v)$ is the vacuum energy.
Substituting eqs. \ref{ansatz} and \ref{potential} in
eq. \ref{lagrangian1},
\begin{eqnarray}
\label{lagrangianA}
L = \left(\frac{\pi}{2}\right)^{\frac{d}{2}}R^d\left[\frac{1}{2}\dot A^2 -
V(A)\right],
\end{eqnarray}
\noindent
with
\begin{equation}
\label{VA}
V(A) = \frac{1}{2}\omega_0^2A^2 + \sum_{n=3}^{4}
\left(\frac{2}{n}\right)^{d/2}\frac{1}{n!}
V^n(\phi_v)A^n, 
\end{equation}
\noindent
where $V^n(\phi_v)\equiv \partial^n  V(\phi_v)/\partial \phi^n$ and
$\omega_{0}^2\equiv V''(\phi_v)+d/R^2$ is the linear frequency. (Primes will
also denote partial derivatives with respect to $\phi$.) For potentials with
$V''(\phi_v)>0$, $V(A)$ is the potential energy of a nonlinear
oscillator with at least one equilibrium point (at $A=0$) for any value of $R$. 
For
conservative motion we can write the energy as,
\begin{eqnarray}
\label{energy}
E = \left(\frac{\pi}{2}\right)^{\frac{d}{2}}R^d\left[\frac{1}{2}\dot A^2 +
V(A)\right] = \left(\frac{\pi}{2}\right)^{\frac{d}{2}}R^d V(A_{\rm max}),
\end{eqnarray}
where $A_{\rm max}$ is
the positive turning point ($A_{\rm max} >$ 0) of $A(t)$. 

We also write the expression for the frequency $\omega$ associated with the
conservative motion between the two turning points,
\begin{equation}
\label{frequency}
\frac{2\pi}{\omega}=T=\int_{0}^T dt=2\int_{A_{\rm min}}^{A_{\rm max}}
\frac{dA}{\dot A},
\end{equation}
where $T$ is the period of oscillation, $\dot A =
\left[2E/c_R-2V(A)\right]^{\frac{1}{2}}$, $c_R\equiv (\pi/2)^{d/2} R^d$, and
$A_{\rm min}$ is related to $A_{\rm max}$ by $V(A_{\rm min}) =
V(A_{\rm max})$.

It was noted in \cite{copeland, dosc} that the location of $A_{\rm max}$
relative to the inflection points of $V(A)$ is an accurate indicator of oscillon
existence.  This can be understood by noting that since the inflection points
separate regions of opposite curvature in $V(A)$, the particular region probed
by $A_{\rm max}$ indicates the degree to which the field configuration experiences the stabilizing
nonlinearities of the potential $V(\phi)$.  Following
\cite{dosc}, we therefore adopt $V''(A_{\rm max})$ as an indicator of oscillon
stability: increasingly negative(positive) values lead to increased
stability(instability).  Defining $I(A_{\rm max},R)\equiv V''(A_{\rm max})$,
\begin{equation}
\label{stability}
I(A_{\rm max},R) = \omega_{0}^2 +  \sum_{n=3}^{4}
\left(\frac{2}{n}\right)^{d/2}\frac{1}{(n-2)!}
V^n(\phi_v)A^{n-2}_{\rm max}.
\end{equation}
In figure \ref{fig:energyI} we show the (parabolic) level curve corresponding to
$I(A_{\rm max}, R)=0$.  Since values inside (outside) the parabola have $I<0$
$(I>0)$, it is easy to show that, for a given $R$, there exists a stable region
provided that $R > R_{\rm min}$ where \cite{dosc}
\begin{equation}
\label{minimum-radius}
R_{\rm min}^2 = d\left[\frac{1}{2}\left(\frac{2^{3/2}}{3}\right)^d
\frac{(V''')^2}{V^{IV}}-V''\right]^{-1}.
\end{equation}
As the figure indicates, when $R=R_{\rm min}$, there is only one value of
$A_{\rm max}$ located in the stable region, given by ${\bar
A}_0=-(4/3)^{d/2}V'''/V^{IV}$.

\section{Life Story of An Oscillon}

\begin{figure}
\includegraphics[width=3in,height=3in]{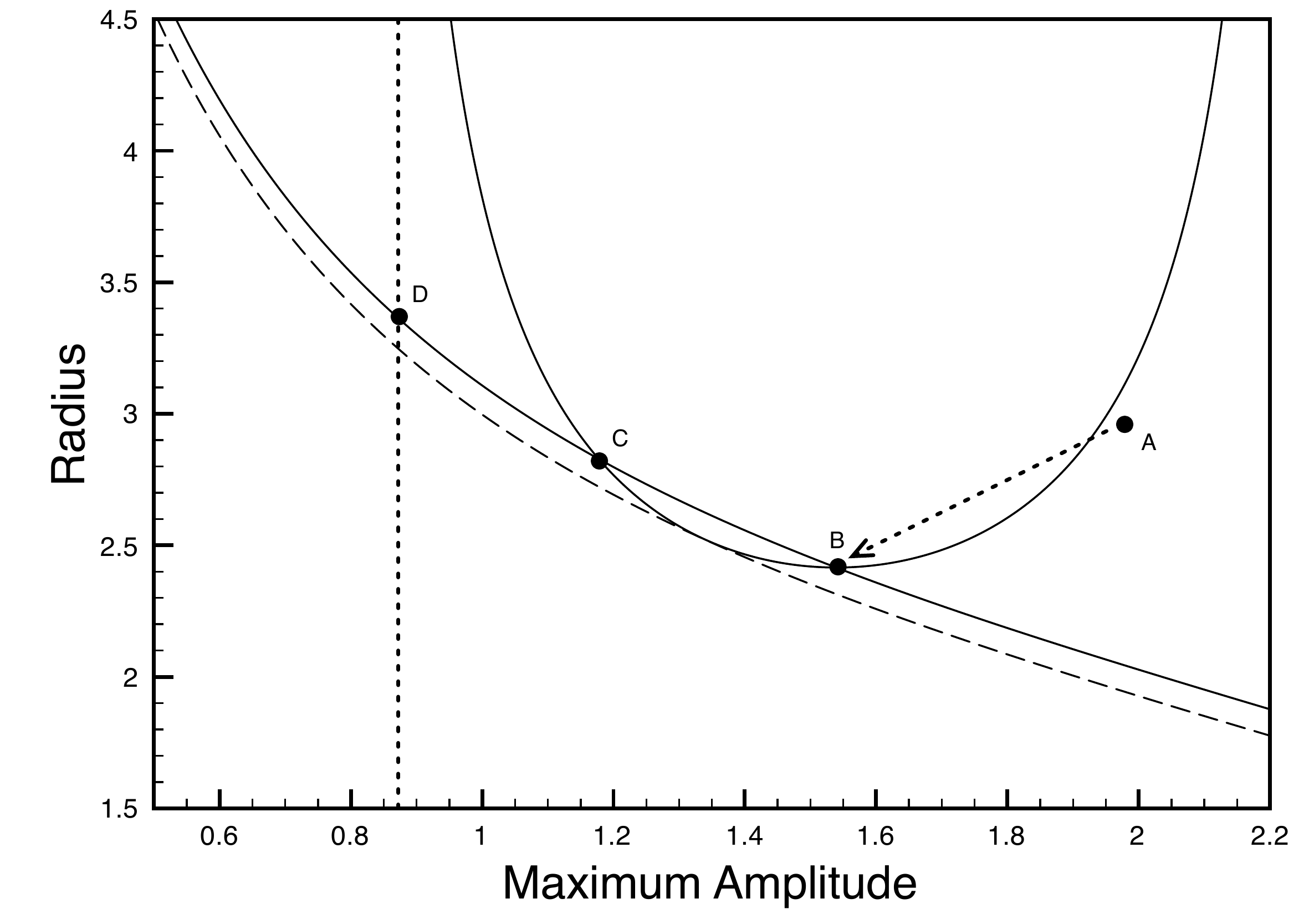}
\renewcommand{\baselinestretch}{1.0}
\caption[EandI]
{Two curves of constant energy, $E_{\rm osc} = 41.3$ (continuous line), and $
E_{\rm attract} = 37.7$ (dashed line), together with the parabolic level curve $I(A_{\rm
max}, R)=0$.  Vertical dashed line locates the asymptote of $I=0$.}
\label{fig:energyI}
\end{figure}

In figure \ref{fig:energyI}, the continuous and dashed lines show curves of
constant energy (eq. \ref{energy}) as a
function of $R$ and $A_{\rm max}$ for a symmetric double-well potential,
\begin{equation}
\label{examplepot}
V(\phi) = \phi^2 - \phi^3 +\phi^4/4,
\end{equation}
\noindent
that is, $g_1=0$, $g_2=2$, $g_3=-6$, $g_4=6$, $\phi_v=0$ in eq. \ref{VA}.
[Quantities were made dimensionless with $x^{\mu} =
x'^{\mu}(\sqrt{g^4}\phi_v)^{-1}$ and the primes have been suppressed.]
We also plot the parabola $I(R,A_{\rm max}) =0$ marking the boundary between the
stable region ($I<0$, above the curve) and the unstable region ($I>0$, below the
curve).  Note that the presence of an oscillon in the unstable region does not
necessarily mean that it will immediately decay; the degree of stability is
given roughly by the vertical distance of a point to the curve $I=0$.

\begin{figure}[t]
\includegraphics[width=3in,height=3in]{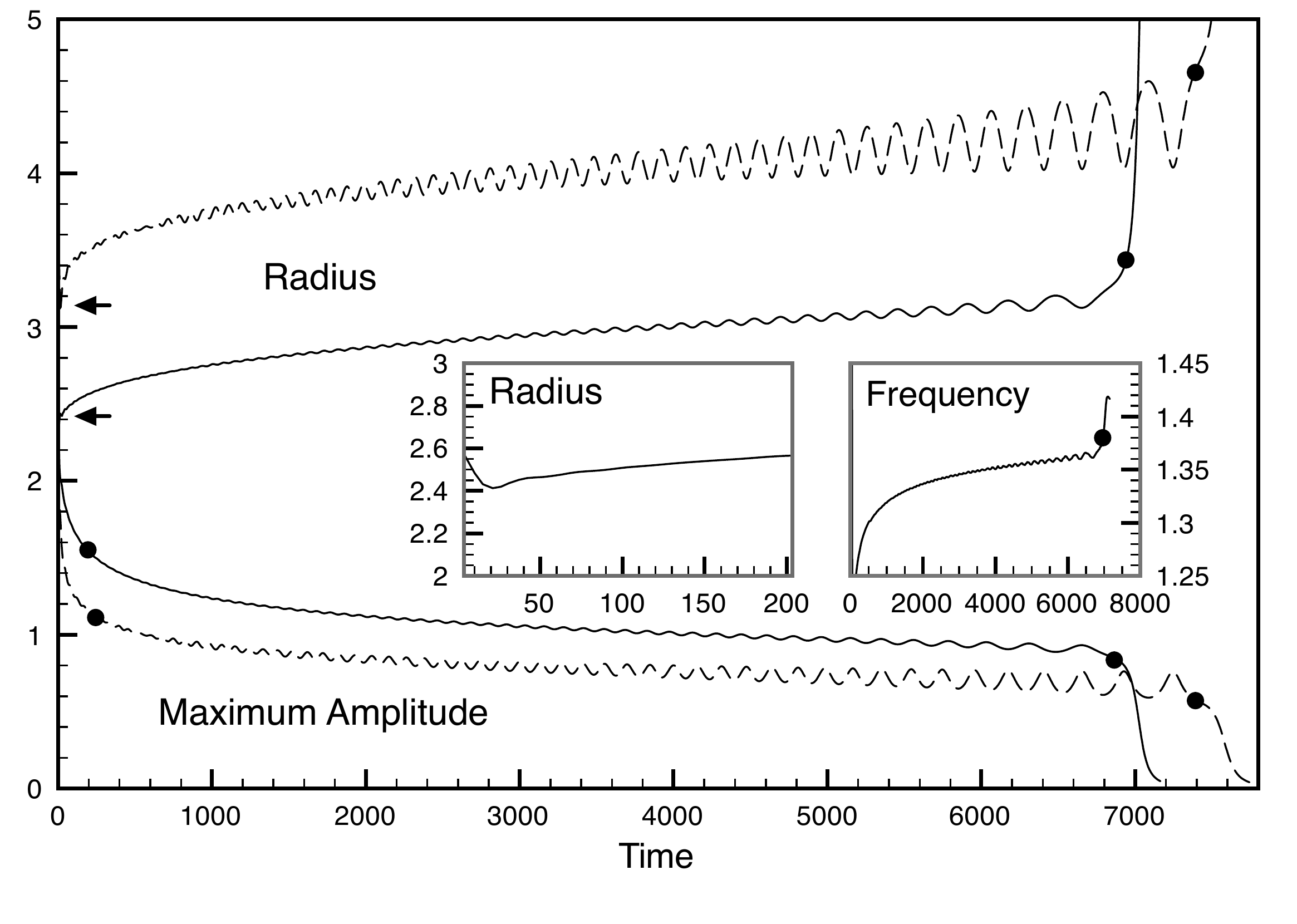}
\renewcommand{\baselinestretch}{1.0}
\caption[R for oscillons]
{The maximum amplitude and radius as a function of time for a configuration with
$R_0=2.86$ and $A_0=2$ in the symmetric double well of eq. \ref{examplepot}
(continuous lines), and for an asymmetric double well
[$V(\phi)=\frac{1}{2}\phi^2-\frac{2.16}{3}\phi^3+\frac{1}{4}\phi^4$] with $A_{0}
= 2$ and $R_{0} = 4$ (dashed lines). The analytical predictions at the points
$(A_B, R_B)$ and $(A_D,R_D)$ are indicated by arrows and dots. The insets show
the minimum radius and frequency for the symmetric double well.}
\label{fig:radiusevol}
\end{figure}

Using the curve $I=0$ in fig. \ref{fig:energyI}, we can describe the trajectory
modeling the onset of an oscillon from an initial configuration starting at a somewhat arbitrary point $A$.  On its way to an oscillon, the configuration will
radiate excess energy until reaching point $B$. 
To determine the location of point $B$, note that the onset of the oscillon
phase is marked by a minimum radius. (In what follows it will be
clear that this \textit{must} be true if radiation radically slows down.) 
Taking the minimum radius given by eq. \ref{minimum-radius}, and using that
the oscillon stage should begin in the stable region, fixes
the location of point B to $A={\bar A}_0$ and $R=R_{\rm min}$, as shown in
figure \ref{fig:energyI}. As we shall see, this is confirmed by numerical results.
For the potential of eq. \ref{examplepot}, $R_{\rm min}=2.42$ and $A_B={\bar
A}_0=(4/3)^{3/2}=1.54$, in excellent agreement with the inset in figure
\ref{fig:radiusevol}.

Since our model approximates an oscillon as having constant energy, we input these values
into eq. \ref{energy}, to obtain the oscillon energy, $E_{\rm osc}$:
\begin{eqnarray}
\label{Eplateau}
E_{\rm osc} \simeq E(A_B,R_B) = \left(\frac{\pi}{2}\right)^{d/2}R_B^dV(A_B, R_B)
= \\ \nonumber
\frac{1}{8} \left(\frac{\pi}{2}\right)^{d/2}\left (\frac{8\sqrt{2}}{9}\right
)^d\frac{(V''')^4}{(V^{IV})^3}R_B^d=
41.3.
\end{eqnarray}
where the numerical value given is for $d=3$ and the potential of eq.
\ref{examplepot}. Compare with numerical results \cite{gleiser,copeland}, where,
roughly, $42\leq E_{\rm osc}\leq 45$ during the oscillon's lifetime.

In addition to the above estimate for $E_{\rm osc}$, we can derive an absolute
lower bound for the oscillon energy.  This is found by noting that
the dashed line in fig. \ref{fig:energyI} shows a critical energy,
denoted $E_{\rm attract}$, below which the energy curves do not probe the
stable region.  Using the $I=0$ condition, one can eliminate $R$ from the expression of the energy (eq. \ref{energy}), which becomes a function only of $A_{\rm max}$. Remarkably, this function has a minimum, the asymptote energy of figure \ref{fig:energyI}.
For the potential of
eq. \ref{examplepot}, $E_{\rm attract} = 37.7$ ($d=3$) and $E_{\rm attract} = 4.44$ ($d=2$).

Having arrived at point $B$ (after radiating excess initial energy) the oscillon
proceeds leftward along its line of constant energy, $E_{\rm
osc}$ (Traveling rightward would cause $R$ to tend to zero, which
is unphysical).  When the oscillon passes point $C$, its distance from
$I=0$ begins to increase and its instability grows.  To
estimate the point of oscillon decay we note that, at point $D$, the vertical
distance to the $I=0$ curve becomes infinite.  Mathematically, point $D$ is found by taking the
limit $R\rightarrow \infty$ in the expression $I=0$ and solving for $A_{D}$, the
decay amplitude. For a cubic potential this gives, $A_D=-(3/2)^{3/2}(V''/V''')$.
For a quartic potential,
\begin{equation}
\label{AD}
A_D = A_B\left [1 - \left (1 - 2\left (\frac{3\sqrt{2}}{4}\right
)^d\frac{V^{IV}V''}{V'''^2}\right )^{1/2} \right ]
\simeq 0.84,
\end{equation}
where $A_B$ was obtained above and the numerical result is for the potential of
eq. \ref{examplepot} in $d=3$.

To obtain the final radius $R_D$, we evaluate eq. \ref{Eplateau} at point $D$:
$E_{\rm osc} = (\pi/2)^{d/2}R_D^dV(A_D,R_D)$. Solving for $R_D$,
\begin{equation}
\label{RD}
g(A_D)R_D^d + \left (\frac{d}{2}A_D^2\right )R_D^{d-2} - E_{\rm osc}\left
(\frac{2}{\pi}\right )^{d/2}=0,
\end{equation}
where $g(A_D)\equiv V''A_D^2/2 +\sum_{n=3}^4(2/n)^{d/2}V^nA_D^n/n!$. In $d=3$,
one gets a cubic equation for $R_D$. For the potential of eq. \ref{examplepot},
with
$E_{\rm osc}=41.3$ and $A_D=0.84$, we obtain, $R_D=3.43$. In figure
\ref{fig:radiusevol}, we compare numerical and analytical values for $(A_B,R_B)$
(left arrows and dots) and $(A_D,R_D)$ (right dots) for this potential
(continuous lines) and for an asymmetric potential (dashed lines). [The numerical simulation measures $R$ by fitting a Gaussian to the oscillon and reading off the radius.]

Since we know $A_{\rm max}(t)$ throughout the oscillon's life, we can use the
potential $V(A)$ to compute $A_{\rm min}$ and then use eq. \ref{frequency} to
obtain $\omega(t)$.  For example, using $A_{\rm max}=A_D$ gives the decay (or
critical) frequency, $\omega_D=\omega_{\rm crit}=1.38< \omega_{\rm
mass}=V''(\phi_v)=\sqrt{2}$. In the inset of figure \ref{fig:radiusevol} we compare
analytical (dot) and numerical values.  It should be noted that this model
correctly predicts that the oscillon frequency tends to a maximum (as observed
in, for example, \cite{saffian}) as it approaches decay.

\section{Oscillon Lifetime}

Due to the remarkable near-periodicity of an oscillon, its spectral function in
frequency space displays a series of narrow peaks, with the dominant one just
below the mass frequency \cite{hindmarsh1}. Only a small radiating tail
penetrates the region with $\omega > \omega_{\rm mass}$.  As in ref.
\cite{hindmarsh1}, we model the power spectrum of the oscillon core amplitude
$\phi(0, t)$ around its dominant frequency peak with a Breit-Wigner profile. 
Since the power is proportional to the square
of the amplitude of a given frequency component, we write
\begin{equation}
\label{lorentz}
a^{2}(\omega) =K \left [\left(\frac{\omega-\omega_{\rm
osc}}{\frac{1}{2}\Gamma}\right)^2 + 1\right ]^{-1},
\end{equation}
where $a^{2}(\omega)$ is the square of the amplitude of the frequency component
of $\phi(0, t)$ with frequency $\omega$, and $K$ is a constant to be determined.  Given that the amplitude of this primary
frequency peak is approximately that of the oscillon itself (because the first
peak is narrow and holds most of the power) $a^{2}(\omega_{\rm osc}) = K \approx
A_{\rm max}^{2}$, the oscillon's core amplitude.  Since the constant $\Gamma$
setting the width of the resonance peak is related to the (inverse) timescale
associated with the instantaneous radiation rate of the oscillon, it is natural
to let $\Gamma = \frac{d \dot E}{dE}$, where $E$ is the oscillon energy.

Most of the (small-amplitude) radiation leaking from the oscillon is contained
in a mode with wavelength commensurate with the size of the oscillon -- that is, with
wavelength $\sim 4R$.  The frequency of this wave is therefore $\omega_{\rm rad}
= \sqrt{\omega_{\rm mass}^{2} + (2 \pi/4R)^2}$.

Now consider an infinitesimally thin shell of radius $R$ around the oscillon
with volume $c_{d}R^{d-1}dr$, which is filled with the outgoing radiation wave
($\omega=\omega_{\rm rad}$) traveling with speed $v_{\rm rad}$.  Within this
thin shell, the wave can be approximated as a $1d$ plane wave of constant
amplitude; hence its energy density is $A_{\rm rad}^2 \omega_{\rm rad}^2/2$,
where $A_{\rm rad}$ is the amplitude of the radiation wave in the shell.  Since
$A_{\rm rad}^2$ is given by $a^2(\omega_{\rm rad})$, the total radiation energy
contained in the thin shell is $a^{2}(\omega_{\rm rad})
\frac{\omega_{\rm rad}^2}{2} c_{d}R^{d-1} dr$.  The radiation
travels outward at a speed $v_{\rm rad} = \omega_{\rm rad}/k_{\rm rad}$, and $dr
= v_{\rm rad} dt$.  The amount of energy lost by the oscillon per unit time,
$- \dot E$, is therefore
\begin{equation}
\label{final}
-\dot E = A^2\frac{1}{2}\omega_{\rm rad}^2 c_{d} R^{d-1}
v_{\rm rad}\left [\left(\frac{\omega_{\rm rad}-\omega_{\rm
osc}}{\frac{1}{2}\frac{d \dot
E}{dE}}\right)^{2} + 1\right ]^{-1}.
\end{equation}
If the oscillon is long lived, $\omega_{\rm rad}-\omega_{\rm osc} \gg d \dot
E/dE$. 
In this case, and writing $\frac{d \dot E}{dE} = \frac{d \dot
E}{dt}\frac{1}{\dot E}$, we have
\begin{equation}
\label{radequation}
\ddot E^{2} + 4\alpha \dot E^{3} = 0,
\end{equation}
where $ \alpha \equiv
2(1-\omega_{\rm osc}/\omega_{\rm rad})^{2}[A^2 c_{d}
R^{d-1}v_{\rm rad}]^{-1}.$

To integrate eq.
\ref{radequation},  we take $\alpha$ to be constant, a good approximation for an
oscillon. Then,
\begin{equation}
\label{Eoft}
E(t) = \frac{E_{i} - E_{\infty}}{\alpha(E_{i} - E_{\infty})t + 1}+ E_{\infty},
\end{equation}
where $E_{\infty}$ is the asymptotic energy as $t\rightarrow \infty$ and $E_i$ the initial energy. 
The oscillon decays at energy $E_{D} > E_{\infty}$ in a time
$\tau_{\rm life}$ given by
\begin{equation}
\label{life}
\tau_{\rm life} = \frac{1}{\alpha}\left(\frac{1}{E_{D} - E_{\infty}} -
\frac{1}{E_{i} -
E_{\infty}}\right). 
\end{equation} 
If $E_i \sim E_{D}$, the lifetime will be approximately
zero, as expected.  When $E_i \gg E_{D}$, the
lifetime becomes independent of $E_{i}$ and tends to a maximum,  $\tau_{\rm
maxlife}\simeq \frac{1}{\alpha}\frac{1}{E_{D} - E_{\infty}}$.
To evaluate this expression, note that $E_{D}$ is simply the plateau energy
obtained in eq. \ref{Eplateau}.  For $E_{\infty}$, being the minimum energy the
oscillon may possess, it is most natural to choose $E_{\infty} = E_{\rm
attract}$.  For the other parameters in $\alpha$, we analytically
calculate the average of each quantity over the stable phase of
the oscillon's life ($C \rightarrow D$ in figure \ref{fig:energyI}).  For
the potential $V(\phi) = \phi^2 - \phi^3 + \frac{1}{4}\phi^4$, we obtain
$\tau_{\rm maxlife}\approx 6300$, which is quite accurate.

Finally, note that eq. \ref{radequation} admits the solution, $\dot E = -\alpha(E-E_{\rm attract})^2$. Thus, if $E\rightarrow E_{\rm attract},~\dot E\rightarrow 0$. In $d=2$,  one can show numerically that the lowest energy oscillon has $E_{\rm osc}=E_{\rm attract}=4.44$. We thus see why $\tau_{\rm maxlife}\rightarrow \infty$ in $d=2$: the oscillon reaches its attractor value and stops radiating.

We have analytically computed all relevant physical properties of scalar field
oscillons in quartic polynomial potentials, confirming that oscillons are indeed
attractors in field configuration space and that their lifetimes can be
estimated by studying their main radiating mode. Our results suggest why $d=2$
oscillons seem to be stable: they reach their attractor state before decaying.
We intend to extend our approach to oscillons in models with gauge fields and
with more complicated potentials. This work was partially supported by a
National Science Foundation grant PHY-0653341.

 \end{document}